\def\fnote#1{\footnote}

\documentclass[a4paper,12pt]{article}
\usepackage[dvips]{graphics}

\begin{document}
\bibliographystyle{unsrt}
\begin{centering}
{\Large Concentrator of laser energy \\ for thin vapour cloud
production near  a surface} \vspace{0.5 cm}

{\bf P.I.Melnikov\footnote{Corresponding author.
Address: Lavrentyev av. 11, BINP, 630090 Novosibirsk, Russia.
Phone: +7(383)2359 285. Fax: +7(383)235 2163. E-mail:
melnikov@inp.nsk.su}, B.A.Knyazev, J.B.Greenly} \vspace{0.5 cm}

{\small \em Novosibirsk State University, 630090 Novosibirsk,
Russia.}

{\small \em Budker Institute of Nuclear Physics,630090
Novosibirsk, Russia.}

{\small \em Laboratory of Plasma Studies, Cornell University,
Upson Hall 369,}

{\small \em Ithaca, NY 14853.}

\end{centering}

\medskip
----------------------------------------------------------
\begin{abstract}
A novel scheme is presented for production of a thin ($<1$ mm)
uniform vapor layer over a large surface area ($>100$ cm$^2$)
by pulsed laser ablation of a solid surface. Instead of
dispersing the laser energy uniformly over the surface, a
modified Fabry-Perot interferometer is employed to concentrate
the laser energy in very narrow closely-spaced concentric
rings. This approach may be optimized to minimum total laser
energy for the desired vapor density. Furthermore, since the
vapor is produced from a small fraction of the total surface
area, the local ablation depth is large, which minimized the
fraction of surface contamination in the vapor.

{\em Key words: laser evaporation, thin gas layer formation.}
\end{abstract}
-----------------------------------------------------------
\vspace{1cm}

\section{Introduction}

\hspace{30pt}In principle, an ideal plasma-based ion source
for an ion beam would provide a perfectly uniform, very thin
layer of fully ionized plasma with a single ion species, at a
temperature low enough to introduce negligible beam
divergence. For magnetically insulated ion diodes for
light-ion inertial fusion drivers, these criteria in practice
would require a plasma layer less than 1 mm thick, with areal
density greater than 10$^{15}$ cm$^{-2}$, uniform to within
10$\%$ to prevent unacceptable perturbation of the plasma
surface smoothness causing beam divergence, and ion
temperature below 50 eV. Pulsed-laser ablation of vapor from a
surface is one possible means of providing a gas layer to form
such a plasma, before application of the high-voltage pulse.
This gas layer can be ionized either by near-surface discharge
in the applied field, or in advance, by means of
photoresonance laser ionization of the vapor \cite{blustpl97}.
By this technique one could obtain space-charge-limited current
from all the anode plasma surface and, consequently,  to eliminate
additional sources of the ion beam divergence \cite{Slutz}.

Work at Sandia National Laboratories \cite{stibeams90} has
produced anode vapor layers in this way, by dispersing laser
energy over the desired anode surface area. High uniformity
and purity is very difficult to achieve by this method,
especially since the areal density of vapor desired is rather
small, but the area is large ($>100$ cm$^2$), requiring the
laser energy to be dispersed uniformly to an energy  density
not far above the threshold for ablation, a regime in which
vapor production is a strong function of laser energy density
and surface contaminants may dominate vapor production.

In this paper we investigate an alternative scheme in which
the laser is intentionally concentrated into a series of
concentric narrow rings. Vapor is produced in these rings,
which are closely enough spaced ($\sim 0.5$ mm) that the
ablated clouds merge to an adequately uniform layer of 0.5 mm
thickness. The concentration allows the evaporation to be done
at a higher, optimum power density for a particular laser and
anode material, so that the overall laser energy is minimized
(see \cite{knyltpp79,knysjtp86}). In addition, for a given
total amount of vapor, the depth of ablation in the rings is
larger than for uniform illumination by the ratio of total
area to ring area, so surface contaminants to bulk material ratio could be
much less in comparison with a case of the disperse  radiation distribution
at a lower power density.

\section{Principle of operation}

\begin{figure}[t]
\centering
\unitlength = 1 cm
\resizebox{10 cm}{!}{\includegraphics{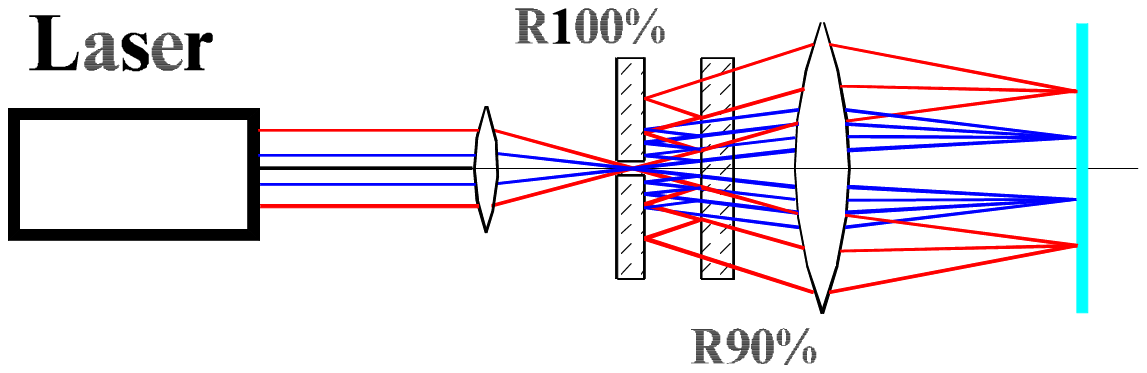}}
\caption[Scheme of the concentrator.]
{Scheme of the concentrator.}
\label{f1}
\end{figure}

\hspace{30pt}The scheme of the concentrator  is shown in
Fig.1. The concentrator is based on  a Fabry-Perot
interferometer. A laser beam is introduced by an input
focusing lens into the interferometer through a small hole in
the first mirror. The output lens focuses the light after the
interferometer to produce concentric thin rings at the target
surface. Most of  the energy of the laser concentrates in
these rings on the surface, except only the central part of
the beam which is lost by reflection back through the hole.
The device including the input lens, the interferometer, and
the output lens we will refer to  as a concentrator. The power
density in the narrow rings is chosen to be high enough to
evaporate surface material efficiently. The amount of material
evaporated per unit laser energy peaks at a particular optimum
value. If this value is chosen for the rings of illumination,
a desired area-averaged vapor density is produced with minimum
laser energy.

We will use the following terms: $r_1$, $r_2$ are reflection
coefficients for first and second mirrors ($r_1\simeq 100 \%$,
$r_2 < 100 \%$); $\tau_1$, $\tau_2$ are transmission
coefficients (defined as $\tau=1-r$); $\lambda$ is the
wavelength of laser generation, $\Delta$ is the distance
between mirrors.

Consider a ray of the laser beam focused by the input lens
with intensity $I_0$ on the input of the concentrator and
angle $\theta$ to the axis. The intensity of this ray behind
the back mirror is $\tau_2I_0$. The intensity of the reflected
part of the ray after the first reflection from the front
mirror and passing through the back mirror is
$r_2r_1\tau_2I_0$, and after $k$th reflection --
$(r_2r_1)^{k}\tau_2I_0$. These rays interfere at the focal
plane of the output lens. The electric field amplitude of the
$k$th is equal to
\begin{equation}
E_k=\sqrt{8\pi(r_1r_2)^{k}\tau_2I_0}\cdot e^{(k-1)\varphi i} \
,  \ k=0,1,2... \  , \end{equation}
where
\begin{equation}
\varphi=4\pi\frac{\Delta}{\lambda}\cos\theta \ . \label{delta}
\end{equation}
is the phase difference of subsequent rays. The
number of the interfering rays N is limited by the radius R
of the interferometer, and resulting the amplitude on the
target is $E=\sqrt{8\pi \tau_2 I_0}\sum_{k=0}^N
(\sqrt{r_1r_2})^k e^{i k \varphi}$, or
\begin{equation}
E=\sqrt{8\pi \tau_2 I_0}\frac{\bigl (\sqrt{r_1r_2}
e^{i\varphi}\bigr )^{N+1}-1}{\sqrt{r_1r_2}e^{\varphi i}-1} \ ,
\end{equation}
where $N=R/(2\Delta\tan\theta)$.
Consequently the distribution of the intensity at a limited
$N$ is
\begin{equation} I=\frac{EE^*}{8\pi}=\tau_2
I_0\frac{({r_1r_2})^N(r_1r_2-2\sqrt{r_1r_2}\cos{(N+1)\varphi})+
1} {(1-\sqrt{r_1r_2})^2+4\sqrt{r_1r_2}\sin^2 (\varphi/2)}
\ . \end{equation}
For $N$ large enough (for small $\theta$)
this expression reduces to the well known Airy-function
\begin{equation}
I=I_0\frac{\tau_2}{(1-\sqrt{r_2r_2})^2+4\sqrt{r_1r_2}\sin^2
(\varphi/2)}. \label{I} \end{equation}
The intensity $I$ in
maxima increases rapidly with  $r_1$ and $r_2$ close to 1.
For this case $\sqrt{r_2r_1} \simeq
1-(\tau_1+\tau_2)/2$, and
\begin{equation}
I=\frac{I_{max}}{1+(16/(\tau_2+\tau_1)^2)\sin^2(\varphi/2)}
\ , \ \ I_{max}=\frac{4\tau_2 I_0}{(\tau_1+\tau_2)^2}
\ . \label{II} \end{equation}

For the distribution of laser radiation over the target
surface ``intensities'' $I$ and $I_0$ in the above written
formulae have to be replaced by the radiation power densities
$F=dP/dS$ and $F_0=dP_0/dS$, which are functions of radius
$\rho$ and azimuthal angle $\psi$ on the target ($dS=\rho
d\rho d\psi$). In contrast to a conventional Fabry-Perot
interferometer, where the power density cannot be higher then
the initial power density, for the concentrator the power
density in the peaks can be much higher than the incident
power density. For example,  if $\tau_1=0$ and $\tau_2=0.04$,
$F_{max}=100F_0$, {\it e.g.} the concentrator multiplies the
peak power density one hundred times.

There may be substantial advantages of such an energy
distribution for surface gas layer production. The maximum
intensity for all the rings would be equal if the initial
laser beam is uniform, and the averaged power density over any
area of dimension larger than the ring spacing is equal to
 $<F>=F_0$, so that for $F_0=$const the distribution of
 averaged power density from the concentrator would be
 constant. The multiplication factor $K=4/\tau_2$ (if
 $\tau_1=0$) gives not only the power density multiplication
 in the rings but also the ratio of total area to the
 illuminated ring area.

\section{Vapor cloud production: a practical example}

\hspace{30pt}The concentrator described can be applied to thin
vapor cloud production near an anode surface of ion
accelerator. Figure 2 shows an example of using of the
concentrator in a magnetically insulated ion diode. The upper
diagram shows direct illumination of the anode for simplicity,
while the lower diagram shows a way to remove the concentrator
from the ion beam path.

\begin{figure}[t]
\centering
\unitlength = 1 cm
\resizebox{14 cm}{!}{\includegraphics{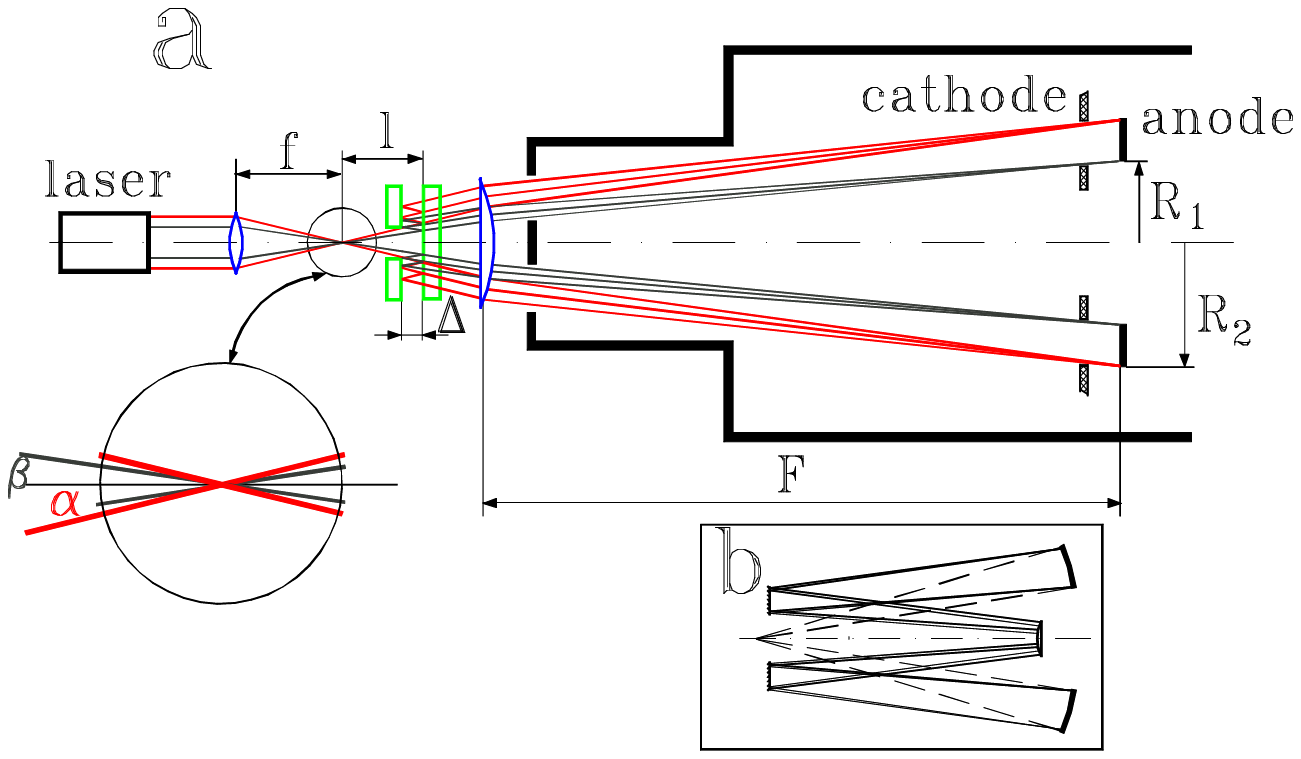}}
\caption{Thin gas layer formation on the anode with the concentrator.}
\label{f2}
\end{figure}

Let us introduce the parameters that must be involved in the
calculation process. $R_1$, $R_2$ are the inner and outer
radii of the anode, $\delta R_2$ is the distance between
illuminated rings on anode surface, $F_{opt}$ is power density
needed to evaporate the anode material. The parameters of the
concentrator are: $f_1$, $f_2$, the foci of the input and
output lenses of the concentrator; $\alpha$, $\beta$, the
angles of the input rays that are directed by the output lens
to the  $R_2$ and $R_1$ radii;  $R$ the radius of the
interferometer; $\Delta$, the distance between the mirrors of
the interferometer; $l$, the distance between the output
mirror and the focus of the input lens; $h$, the radius of the
hole; $K$ the multiplying factor; $P$, the  energy density on
the output mirror surface. The laser parameters are:
$\lambda$, the generation wavelength; $\delta \lambda$, the
bandwidth of laser generation; $t$, the pulse duration; $Q$,
the pulse energy.

We now estimate the parameters needed for realizing the
scheme. We choose the COBRA ion diode \cite{COBRA} as an
example. For this diode, $R_1=7.2$ cm, $R_2=9.5$ cm. We desire
$\delta R_2=0.05$ cm, for adequate uniformity of the vapor
layer on a distance $0.05$ cm from the anode surface. For
evaporation by 10 ns laser pulse of an aluminium target we choose
$F_{max}=F_{opt}^{Al} \simeq 230$ MW/cm$^2$ \cite{knypre80},  for
optimum efficiency of vapor production. Experiments on vapor cloud
production (see \cite{Kny-Heid-inv}) show that for every target there
is a range near an optimal power density where it is possible to obtain
low-ionized cold vapors without the laser-induced breakdown.
 It is significant also that
near this value of power density the amount of evaporated
material and the velocity of vapor boundary are only weakly
dependent on power density \cite{basspjetp67}. Thus the vapor
layer could be more uniform than the initial laser power
density $F_0$. We will use the value $f_2=1.5$ m for the
output lens focus, which would work in the configuration of
Fig.2a for the present COBRA diode. Fig.2b shows a possible
arrangement in a focusing diode geometry. The value of $f_2$
may be adjusted to avoid interference of optical elements with
the ion beam. We assume a ruby laser with wavelength $694.3$
nm. For definiteness we put $K=100$, but this figure might be
varied.

The value $\delta R_2$ defines the distance between mirrors.
Actually, the maxima of power density (\ref{II}) correspond to
the zeros of $\sin(\varphi/2)$. So two nearby maxima differ by
angle $\delta \theta$ when $(\varphi_1-\varphi_2)/2=\pi$
\begin{eqnarray} \pi=2\pi
\frac{\Delta}{\lambda}(\cos\theta_1-\cos\theta_2)=2\pi
\frac{\Delta}{\lambda} \theta \delta \theta \ , \\ \delta
R=\delta \theta f_2 =\frac{\lambda}{2\Delta}
\frac{f_2}{\theta}= \frac{\lambda}{2\Delta} \frac{f_2^2}{R} \
. \nonumber \end{eqnarray}
The distance between mirrors is
\begin{equation} \Delta=\frac{\lambda}{\delta
R_2}\frac{f_2^2}{2R_2}=1.65 \ \mbox{cm} \ . \label{D}
\end{equation}
The angles $\alpha$ and $\beta$ are related to
the $R_2$ and $R_1$ by $\alpha={R_2}/{f_2}=6.3\cdot
10^{-2} \ , \ \ \beta={R_1}/{f_2}=4.8\cdot 10^{-2}$. The
distance between input lens focus point and output mirror
surface may now be defined
\begin{equation}
l=\frac{R_2+R_1}{R_2-R_1}\Delta=7.26\cdot \Delta=12 \
\mbox{cm} \ , \label{l} \end{equation} and the hole radius is
\begin{equation} h=(l-\Delta)\alpha=0.65 \ \mbox{cm} \ ,
\label{h}
\end{equation}
The required radius of the
interferometer  can be determined as the radius at which the
laser beam with  the biggest angle $\alpha$ is attenuated by a
factor of e$^{-2}$.
\begin{equation}
R=K\frac{R_2}{f_2}\Delta=10.5 \ \mbox{cm} \ . \label{R}
\end{equation}
The number of interfered rays is
$N=2/\tau_2=K/2=50$ for this interferometer radius.

We estimate the bandwidth of laser generation that is
allowable to provide such narrow rings. The bandwidth must not
lead to widening of the ring more than the width of the half
height of the power density. From (\ref{delta}) and (\ref{II}) we get
\begin{eqnarray}
\frac{\tau_1+\tau_2}{2}& >& 4\pi \Delta
\frac{\delta \lambda}{\lambda^2} \ ,
\nonumber \\
\frac{\delta \lambda}{\lambda} &<& \frac{(\tau_1+\tau_2) \,
\lambda}{8\pi
\Delta}=\frac{\lambda}{2\pi K \Delta}=6.7\cdot 10^{-8} \
,\label{dl} \\
\delta \lambda &<& 4.7\cdot 10^{-5} \ \mbox{nm}
\ . \nonumber
\end{eqnarray}
The last value is close to a theoretical limit and impose strong
requirement on the laser spectrum.

The main critical parameter of the concentrator is the energy
density in the center of the output mirror surface. The mirror
must be of high quality to be not destroyed by the intense
laser beam. \begin{equation} P=F_{opt}^{Al}t\Biggl
(\frac{f_2}{l}\Biggr )^2 K^{-1}=3.6 \ \mbox{J/cm}^2 \ .
\label{P} \end{equation} This value is not too high, but the
local energy density can be  3 times more due to interference
effect. So the mirror must withstand the energy density of 11
J/cm$^2$. Mirrors of such quality are available
\cite{fouspie91}.

The total energy of laser can be calculated from the formula
$Q=F_{opt}^{Al}t\pi R_2^2 K^{-1}=6.5$ J. Only 40$\%$ of the
energy is used in the evaporation, because of
noncorrespondence of the solid circular laser beam and the
hollow annular anode surface. It is more efficient to use an
annular laser beam. A beam with such a structure can be
produced by a Nd laser with nonstable resonator. Use of 1.06
$\mu$m light leads to an increase of  concentrator size (see
(\ref{D}, \ref{l},  \ref{h}, \ref{R})): $\Delta=2.52$ cm,
$l=7.26 \cdot \Delta=18.3$ cm, $R=16.0$ cm, $h=1$ cm. But the
energy density on the output mirror surface would decrease
(see (\ref{P})) $P=1.6$ J/cm$^2$.

Using the 2nd harmonic of Nd laser generation gives twice
smaller size (see (\ref{D}, \ref{l},  \ref{h}, \ref{R})):
$\Delta=1.26$ cm, $l=7.26 \cdot \Delta=9.15$ cm, $R=8$ cm,
$h=0.5$ cm. However, the energy density becomes four times
higher (see (\ref{P})), and reaches a rather high value
$P=6.4$ J/cm$^2$. To use the concentrator in such a regime
would require a mirror that would not damaged by the energy
density $3P=20$ J/cm$^2$. This value is over the damage
threshold for the best mirror coatings \cite{fouspie91}. But
it is possible that the mirror coating can withstand this
energy density because averaged energy density is less 12
J/cm$^2$, and the value above damage threshold is reached
only in very thin rings of  $20\lambda$ of thickness that
would provide fast heat diffusion over the surface due to
thermal conductivity.

\section{Conclusion}

\hspace{30pt}The production of a very uniform, thin vapor
layer above a solid surface by laser ablation is a difficult
problem. We have suggested a method of distributing laser
energy in a series of narrow, high intensity rings that could
be advantageous in producing vapor efficiently, and,
probably, with
minimum sensitivity to surface contaminants and laser
intensity variation. If this concentrated vapor production can
be allowed to expand and merge into a uniform layer, these
advantages might be realized. We leave aside of this paper
consideration of the problem of possible small-scale non-uniformity
of the gas layer produced with the concentrator as well
possible anode plasma instability. Obviously, at first the technique
suggested has to be verified experimentally, and the anode layer
parameters must be measured.

\section{Acknowledgments}

\hspace{30pt}This work was supported, probably in part, by
U.S. Civilian Research and Development Foundation, Award
RP1-239, and Russian Ministry of General and Professional
Education.

\end{document}